\documentclass[twocolumn]{article} 

\usepackage{graphicx}
\usepackage{hyperref}
\usepackage{twoopt}
\usepackage{authblk}
\newcommand{\ackname}{Acknowledgements}
\usepackage{eurosym}
\usepackage{amstext} 
\DeclareRobustCommand{\officialeuro}{%
  \ifmmode\expandafter\text\fi
  {\fontencoding{U}\fontfamily{eurosym}\selectfont e}}

\begin{document}

\title{
\vspace{-1cm}
Considerations for a new low-/moderate-resolution optical 
facility spectrograph at the VLT Coud\'e focus}

\author[1]{Valentin D. Ivanov}
\author[1]{Jean-Louis Lizon}
\affil[1]{European Southern Observatory, Karl-Schwarzschild-Str. 2,
85748 Garching bei M\"unchen, Germany; vivanov@eso.org}
\date{}

\maketitle

\begin{abstract}
{\small
\vspace{-2mm}
Observing at the VLT Coud\'e focus can boost the collecting 
area by combining light from multiple VLT unit telescopes 
(UTs; albeit with some losses in the light train). An 
instrument at the Coud\'e enjoys significant operational 
flexibility advantage: it can be attached to any available 
UT and the ``extra'' instrument can help to match better 
the observing constraints with the current conditions. With 
modifications to the existing train it can even observe in 
parallel with ESPRESSO with different UTs.
}

{\small
Here we consider a general purpose VLT Coud\'e fiber-fed 
low-resolution facility spectrograph -- provisionally named 
{\it Cappuccino} -- suitable for rapid follow up and 
characterization of faint transients, for late-stage 
monitoring of transients and for rapid classification under 
poor conditions or during twilight. Like any instrument of 
its class, {\it Cappuccino} can be used to address diverse 
set of science questions. The building cost can be reduced 
greatly if it is based -- with modest changes -- on existing 
hardware.
}
\end{abstract}

\section{Motivation for an additional low/medium-res
Coud\'e spectrograph at the VLT}

The recent years saw a spiking interest in the time-domain 
astronomy. Sky monitoring and variability has long been a 
mainstay of astronomy, but now we see the community moving 
from a handful of dedicated variability surveys (e.g. EROS, 
MACHO, 2MASSX6 and OGLE) to a number of ambitious and 
efficient wide-field ground- and space-based facilities 
(e.g. VISTA, NGTS, WASP, WASP-S, MASCARA, {\it Kepler/K2}, 
{\it TESS}, etc.). Even more capable facilities are about 
to start operating in the near future (e.g. VCRO). Right 
now the {\it Gaia} mission yields on average 1350 alerts 
for transient events yearly. About 1/3 of them are brighter 
than R$\sim$18\,mag. The expected yield of the {\it Zwicky 
transient survey} (47 sq. deg FoV camera, survey speed of 
over 3000\,deg$^2$ per hour) is a million transients per 
night, and the team behind this project plans a robotized 
spectroscopic follow up of all objects with r$<$18.5\,mag 
with the Palomar 5-m telescope. VCRO will sweep the field 
with over 10 million transients per year. The time scale 
of these events varies in a wide range: from minutes-hours 
for gamma ray bursts, to days-week for kilonovae and to 
many months for some types of supernovae.

The follow up and characterization of this large yield is 
an unprecedented challenge to astronomy. Various research 
teams and community wide working groups -- e.g., 
\cite{2016arXiv161001661N} -- have identified three types 
of necessary follow up facilities:

(1) efficient medium-resolution spectrographs like X-shooter 
and SoXS that cover the (nearly) entire optical/near-infrared 
range and in the infrared they can work between sky lines -- 
aimed mainly, but not exclusively, at the extragalactic 
transients

(2) high miltiplexity facilities that deliver thousands of 
spectra over degree-size fields in moderate/high-resolution 
5000/30000 up to 1.3-1.7\,$\mu$m in the red, similar to the 
incoming 4MOST -- aimed at Milky Way transits that can be 
``injected'' in other on-going surveys

(3) low/moderate-resolution R$\sim$1000-2000 wide wavelength 
coverage (400-1000/1300\,nm) spectrographs, similar to the 
existing combinations FORS2+SINFONI or EFOSC2+SofI -- to 
characterize faint transients, to follow up transients at 
late stages when they have grown faint, or to deliver 
preliminary classification of bright transients thereby 
saving time at other high-pressure instruments

As notes above, ESO offers or will offer soon instruments 
in all these categories, but the third category deserves a 
special discussion. Despite numerous upgrades, SINFONI and 
EFOSC2 are ageing instruments that will be decommissioned 
eventually (albeit SINFONI will be replaced by ERIS). FORS2 
is mounted on UT1 (unit telescope) that is typically 
over-subscribed by a factor of 4 or more, and over the last 
few years UT1 systematically comes as the first or the 
second most demanded among the four VLT units.

Expanding the options that the users have is possible by 
taking advantage of the already-operational Coud\'e facility 
and offering either (i) an additional focus for fiber-fed 
user instrument or by placing there (ii) a new fiber-fed 
low- or moderate-resolution facility spectrograph spanning 
the entire optical range from 360 to $\sim$900\,nm.

Such new instrument will have lower efficiency with respect 
to FORS2 or X-shooter because of the train and fiber losses, 
but will gain over SoXS and 4MOST because of the larger 
collecting area of the 8m class VLT telescope over 4m class 
NTT and VISTA (btw, 4MOST is also fiber-fed with respective 
losses).

A great cost optimization is achievable if the new instrument 
is based on some of the existing non-operational instruments, 
e.g. FORS1, EFOSC2, etc, upgraded with modern filter, grism 
and detector to improve the efficiency. Operationally, the 
instrument can be limited  -- at least initially -- to a 
single mode which means no moving parts. This implies improved 
stability and maintainability, and more straightforward data 
reduction. Other modes can be implemented later, if the 
community requests them.

A fiber-fed Coud\'e spectrograph will potentially suffer from 
poor sky subtraction -- like all single-fiber spectrographs,
because there is no option to monitor the off-target sky 
simultaneously with obtaining the science spectrum. This may
potentially lower the red end of the useful spectral range to
850-900\,nm for faint targets, to exclude regions with strong
sky emission lines.

The Coud\'e facility allows to feed the new instrument with 
light from multiple VLT units, gaining deeper limiting 
magnitude than with a single UT. Another advantage of the 
proposed instrument is the greater operational flexibility -- 
the liberty to be used with any of the UTs. Finally, the new 
instrument will not compete with ESPRESSO, but it will 
complement it because of the incomparable  
resolutions.

\section{Existing infrastructure -- The VLT Coud\'e facility }

The Coud\'e facility includes:

(1) Combined Coud\'e Laboratory (CCL) -- a rectangular room 
of approximately 10.7$\times$20\,m size, partially occupied 
by ESPRESSO \cite{2010SPIE.7735E..0FP}. Vibrations by the 
additional instrument should not disturb ESPRESSO (exhich is
one of the many reasons why an additional Coud\'e instrument 
should not work in the infrared). Additional power and local 
area network connections may be needed.

(2) Coud\'e Room -- a circular room of roughly 8.5-m diameter 
centred on the azimuth axis of the telescope.

(3) Coud\'e Bodega -- located around the Coud\'e room, it was 
used as a warehouse.

(4) Coud\'e Trains -- a number of electric, optical and 
mechanical components, controlled by the telescope control 
software; they transfer the light of the UTs from their 
Nasmyth\,B foci to the CCL; Coud\'e\,A is reserved for VLTI 
and Coud\'e\,B -- for ESPRESSO.

\begin{figure}[!htb]
\centering
\includegraphics[width=8.1cm]{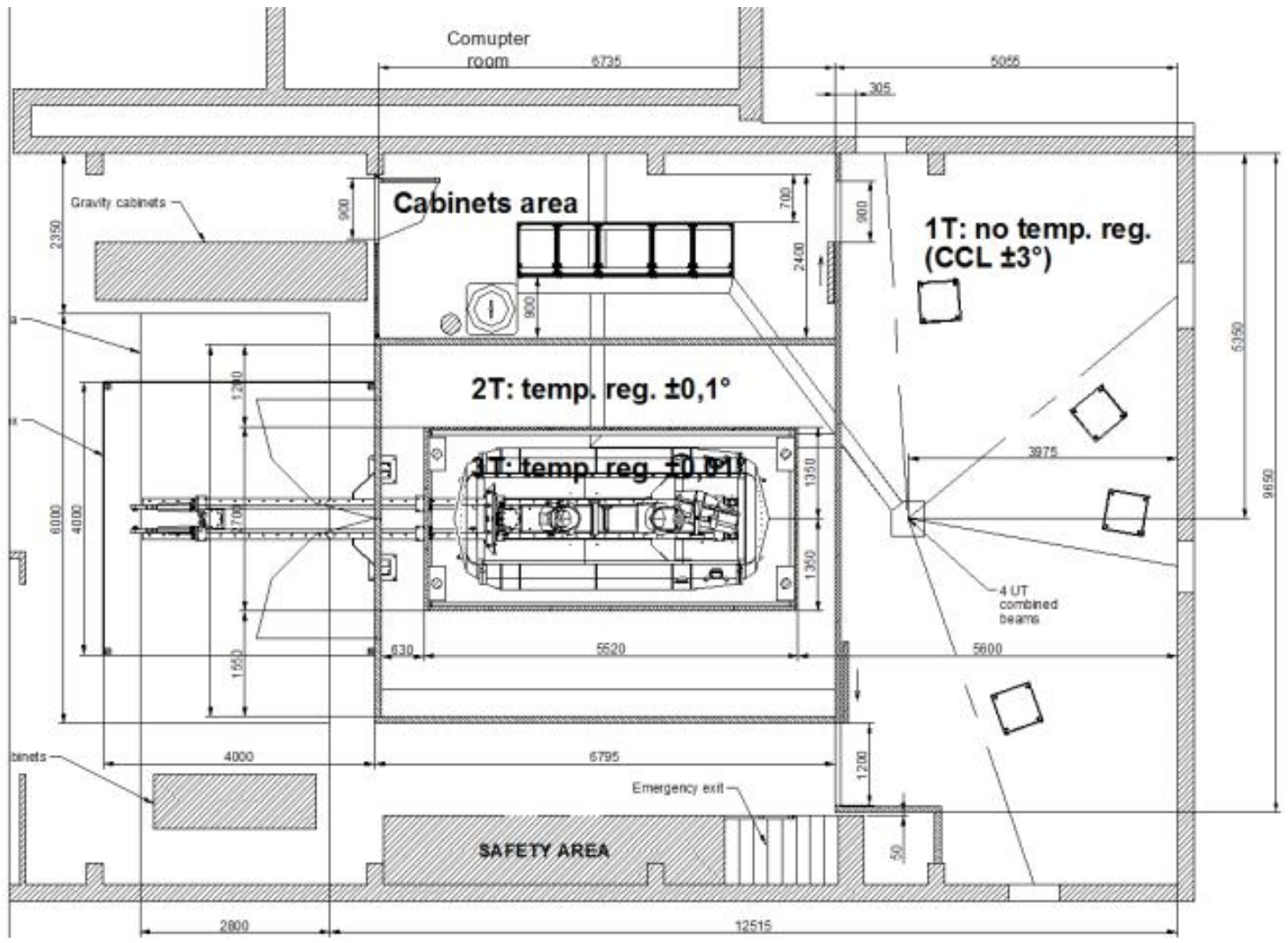}
\caption{CCL, generic view. ESPRESSO is at the centre, the 
light trains from the individual UTs arrive to the CCL on 
the right (UT1, UT2, UT3 and UT4, respectively from top to 
right and to bottom). The compact size and the low weight 
of {\it Cappuccino} allow to place it on a rack above the 
ground, saving space and easing the access to 
it.}\label{fig:CCL}
\end{figure}

These components have existed, to one or another degree, since 
the construction of the VLT but they have been used only 
partially by the VLTI until ESPRESSO entered operations. A 
schematic drawing of the CCL with ESPRESSO is shown in 
Fig.\,\ref{fig:CCL} (reproduced form the ESPRESSO user 
manual\footnote{\protect{\url{https://www.eso.org/sci/facilities/paranal/instruments/espresso/doc.html}}}).

{\it Cappuccino}, the instrument considered here, could be 
located in the CCL next to ESPRESSO. Sharing the CCL with 
ESPRESSO sets limitations on heat production and vibrations, 
but the compact nature of {\it Cappuccino}, its lack of 
moving parts and of closed-cycle coolers promise that these 
limitations will be easily met.

\section{{\it Cappuccino} sub-systems}

{\it Cappuccino} foreseen seen as a simple and manimalistic 
system -- for easy building, operations, maintainence and for 
stability that in turn will make the data processing more
straightforward. The basic components of this instrument are:

- Front end: optics receiving light from the Coud\'e train 
and feeding it into the spectrograph. Additional fiber pick 
ups are glued to the rotating table. In the simplest case it 
is just one pick up, allowing {\it Cappuccino} to use one UT
at a time, in the more advanced case -- four pick ups and a 
beam combiner, allowing it to use multiple UTs. Without 
further modifications {\it Cappuccino} can not work in parallel 
with ESPRESSO, but it is not impossible to image a design 
where the two instruments are fed simultaneously by different 
UTs. The front end performs target acquisition, 
alternates between on-sky observations and the calibration 
unit, and field/pupil stabilization during exposure. 
Similarly to ESPRESSO, it controls the atmospheric dispersion 
correctors (even though they are formally part of the Coud\'e 
train). The new instrument shares the toggling mechanism with 
ESPRESSO.

- Calibration unit: includes lamps for wavelength calibration 
and flat fielding, adjusting their intensity as necessary. It 
is located in the CCL and provides the light via optical fiber.

- Fiber link: connects the front end and the spectrograph. 

- Spectrograph: includes a focal reducer. The slit is formed 
at the end of the fiber link. The dispersing element (prism, 
grism or grating) is mounted on a bench, reducing flexure and
providing sufficient earthquake stability. The light of a 
single order is recorded onto the detector. Finally, there is 
a camera with respective input optics. The spectrograph is 
warm, except for the detector. The detector dewar is equipped 
with a pumping system and all the necessary vacuum and 
temperature sensors.

- Instrument electronics: controls instrument functions and 
telemetry, provides a link to the Instrument workstation. 
Part of it is inside the dewar with the detector.

If the design of the new spectrograph is based on an existing 
instrument, then the new systems (aside from any elements that 
might be upgraded to gain efficiency) are the focal reducer 
and the calibration unit that will feed it via a faiber -- 
because all existing instruments that were considered here 
were directly fed and have f-ratio inconsistent with a fiber; 
their calibrations units typically are designed to shine the 
light on a screen.

\section{{\it Cappuccino} efficiency}

Estimated throughput breakdown (based on the ESPRESSO and 
FORS2 values, whenever possible):

- Coud\'e train: $\geq$80\%

- Front end: $\geq$83\%

- Fiber link: 70\%

- Spectrograph optics: $\sim$70\% at any wavelength (for 
approximately 10 optical elements with $\sim$2\% loss per 
surface)

- Detector: 90\% over most of the wavelength range

A conservative estimate of the total throughput is: 
0.8*0.83*0.7*0.7*0.9$\sim$30\%. As expected, it is lower 
than for FORS2 with its average of 45\% (grism 600I, 
400-1000\,nm range). 4MOST -- another fiber-fed spectrograph -- 
has an observing efficiency of 18\% over 370--950\,nm 
(resolution R$\sim$4000-7500).

\section{{\it Cappuccino} hardware cost}

These cost estimates are for the simplest design case -- a 
single faber pick up allowing {\it Cappuccino} to use one 
UT at a time, and single mode/grating (no moving parts).
Data flow system is not included.

\subsection{Case I: design based on FORS1 or EFOSC2}

This option reuses an existing collimator and camera.
Itemized costs:

(1) Fibre link and feed adaptation: design 0.3 FTE, 
procurement 40\,K\euro

(2) Structure and enclosure for the optic and calibration 
unit: design 0.5 FTE, procurement 60\,K\euro

(3) Grating (grism VPH and mount): design 0.2 FTE, 
procurement 40\,K\euro

(4) Detector system:

(4.1) Cryostat (Cryostat, NGC - new generation controller 
and detector; no new design required, everything exist 
already): test 1 FTE, procurement 55\,K\euro\ (cryostat and 
Cryo vacuum control and equipment), 60\,K\euro\ (NGC), 
150\,K\euro\ (detector)

(4.2) Electronic control (only for calibration source, and 
pick up system ) 30\,K\euro

Total: 2 FTE, 435\,K\euro.

\subsection{Case II: complete new optical system, camera and 
collimator}

This option has the advantage of better optimization, e.g. 
for higher efficiency over wider wavelength range. The 
additional costs over Case I are: for the design 0.5 FTE and 
procurement 300\,K\euro.

Total: 2.5 FTE, 735\,K\euro.

\section{Summary and conclusions}

{\it Cappuccino}, a new general-purpose optical facility 
instrument for the VLT Coud\'e focus is considered. Its main 
advantages are:
\begin{itemize}
\item capability to quickly deliver spectra of faint targets, 
especially ones that rapidly grow faint, e.g. various 
transients;
\item significant improvement of the VLT operational 
flexibility, in effect adding an extra low-res spectrograph 
at each UT that can be used during poor observing conditions;
\item extremely low cost.
\end{itemize}

\section*{\ackname}
This is an extended write up of a poster presented at the 
{\it The Very Large Telescope in 2030 (VLT2030)} ESO workshop, 
held in ESO Garching on June 17-20, 2019. The authors thank 
the organizers for the opportunity to present this idea. We 
also thank Olivier Hainaut, Gerardo \'Avila and Leonardo Vanzi 
for the helpful discussions.


\end{document}